\documentclass[reprint,aps,prl]{revtex4-2}
\usepackage{amssymb,amsmath}
\usepackage{graphicx}
\pdfoutput=1

\usepackage{placeins}
\usepackage{bm}
\usepackage{epstopdf}
\usepackage{color}
\usepackage{subfigure}
\usepackage{parskip}
\usepackage[compact]{titlesec}
\usepackage{hyperref}

\usepackage{esvect}

\definecolor{mycolor}{RGB}{0,30,96}
\hypersetup{
	colorlinks=true,
	allcolors=blue,breaklinks,backref}

\usepackage{cancel}

\usepackage{placeins}

\titlespacing{\section}{0pt}{1em}{0em}
\titlespacing{\subsection}{0pt}{1em}{0em}

\begin{document}

\title{Gain-loss-induced non-Abelian Bloch braids}
\author{Bikashkali Midya}
\email{midya@iiserbpr.ac.in} 
\affiliation{Indian Institute of Science Education and Research Berhampur, Odisha 760003, India}

\begin{abstract}
Onsite gain-loss-induced topological braiding principle of non-Hermitian energy bands is theoretically formulated in multiband lattice models with Hermitian hopping amplitudes. Braid phase transition occurs when the gain-loss parameter is tuned across exceptional point degeneracy. Laboratory realizable effective-Hamiltonians are proposed to realize  braid groups $\mathbb{B}_2$ and $\mathbb{B}_3$ of two and three bands, respectively. While $\mathbb{B}_2$ is trivially Abelian, the group $\mathbb{B}_3$ features  non-Abelian braiding and energy permutation originating from the collective behavior of multiple exceptional points.  Phase diagrams with respect to lattice parameters  to realize braid group generators and their non-commutativity are shown. The proposed theory is conducive to synthesizing exceptional materials for applications in topological  computation and information processing. 
\end{abstract}
\maketitle

 The idea of braids and knots of quantized excitations in certain interacting many-body quantum systems has opened extraordinary possibilities for building blocks of a fault-tolerant topological quantum computer \cite{Kitaev2006,Kitaev2003,Nayak2008}.  While  quantum materials with desired real-space braiding dynamics are notoriously illusive \cite{google2023},  very recently there has been a growing interest in non-trivial knots and braids of energy bands in the reciprocal space of  single-particle non-Hermitian (NH) systems \cite{Ren2013,Hu2021,Wang2021,Patil2022,Yu2022,Woicik2022,Zhang2023,Parto2023,Li2022,Cao2023,Zhang2023b}. Periodic  open systems described by effective NH Hamiltonians generally feature complex band energy manifolds with  exceptional topological properties, not accessible in Hermitian systems, with far-reaching practical consequences \cite{Bergholtz2021,Ding2022,Li2023,Yang2020,Midya2022,Zhao2019,Midya2018,Zhang2022,Weidemann2020,Gong2018,Konotop2018,Gao2023,Shen2018,Xiao2020,Weidemann2022}. Two Bloch bands, for example, of a one-dimensional (1D) NH lattice can wind around, forming a nontrivial topological braid in the $(2+1)-$dimensional space spanned by complex-energy and momentum $(\mbox{Re} E, \mbox{Im} E,k)$ \cite{Wang2021}; this situation is forbidden in a 1D Hermitian lattice.  When lattice parameters are suitably varied,  such NH braid structures undergo a phase transition at the exceptional point (EP) band degeneracy, at which two bands become indistinguishable. Topologically distinct braids of $N$ bands can naturally be classified by the elements of a braid group $\mathbb{B}_N$ such that each braid carries a topological invariance  \cite{Hu2021}. 
 
 Current progress on 1D reciprocal space braiding, however, is limited mainly to lattice models based on the NH coupling principle \cite{Wang2021,Zhang2023,Patil2022,Hu2021,Yu2022}. The possibility of onsite gain-loss-induced braid phase transition and generalization to non-Abelian braiding remain largely unexplored.  Here, we address the following  question : \textit{can onsite gain or loss solely induce braid groups and corresponding topological phase transitions in  an otherwise 1D Hermitian lattice}?  We show by considering toy examples of two and three bands lattices that it is indeed possible to create distinct topological braid elements for the Abelian group $\mathbb{B}_2$ and non-Abelian group $\mathbb{B}_3$, respectively, by judicious choices of externally controllable gain-loss potentials and other Hermitian lattice parameters. Non-Abelian braids under $\mathbb{B}_3$ provide remarkable inequivalent energy permutations when the order in which two braids concatenate is altered. While the handedness of a braid is shown to be tunable by interchanging the roles of gain and loss, the degree of a braid can be tunable by controlling the order of long-range hopping between lattice cells. Parameters regimes to realize elementary braids and their non-commutativity are obtained. Essential topological feature of a braid is determined by the spectral winding number. 
 
 Before describing the gain-loss-induced braiding mechanism, we first brief about general ideas of band spectrum braids.  A braid is a disjoint collection of  interlacing trajectories of $N$ non-degenerate complex energy bands $\{E_1 ,E_2,\cdots,E_N\}$ as $k$ evolves in the Brillouin zone (BZ) $k_0\le k \le k_0+2\pi$; bands are ordered here according to their real parts at an arbitrary $k_0$. This can be mathematically defined by the mapping $E_n(k):BZ\rightarrow \mathbb{C}\times BZ$ for  $n\in\{1,2,\cdots,N\}$;  here, $\mathbb{C}$ represents the complex energy plane. Every braid in $\mathbb{B}_N$ can be obtained as a product of $(N-1)$ generators $\tau_1,\tau_2,\cdots,\tau_{N-1}$ and their inverses $\tau_1^{-1},\tau_2^{-1},\cdots,\tau_{N-1}^{-1}$. These elementary braids $\tau_n$ or $\tau_n^{-1}$ consist of only a single twist between neighboring bands $E_n$ and  $E_{n+1}$ in the positive or negative direction. The identity braid with no twist is denoted here by $\tau^0$.  Collection of braids forms a braid group $\mathbb{B}_N$ under concatenation (the braid $\tau_n\tau_\ell$ is obtained by joining end points of $\tau_n$ with initials of $\tau_\ell$), such that $\tau_n\tau_\ell=\tau_\ell\tau_n$ for $|n-\ell|>1$ and $\tau_n\tau_{n+1}\ne\tau_{n+1}\tau_n$ \cite{Kauffman2013}. The braid group formed by $N>2$ bands is thus non-Abelian. A link or knot is formed when a braid is closed. 
 
 \begin{figure*}[htb!]
 	\centering{
 		\includegraphics[width=0.8\textwidth]{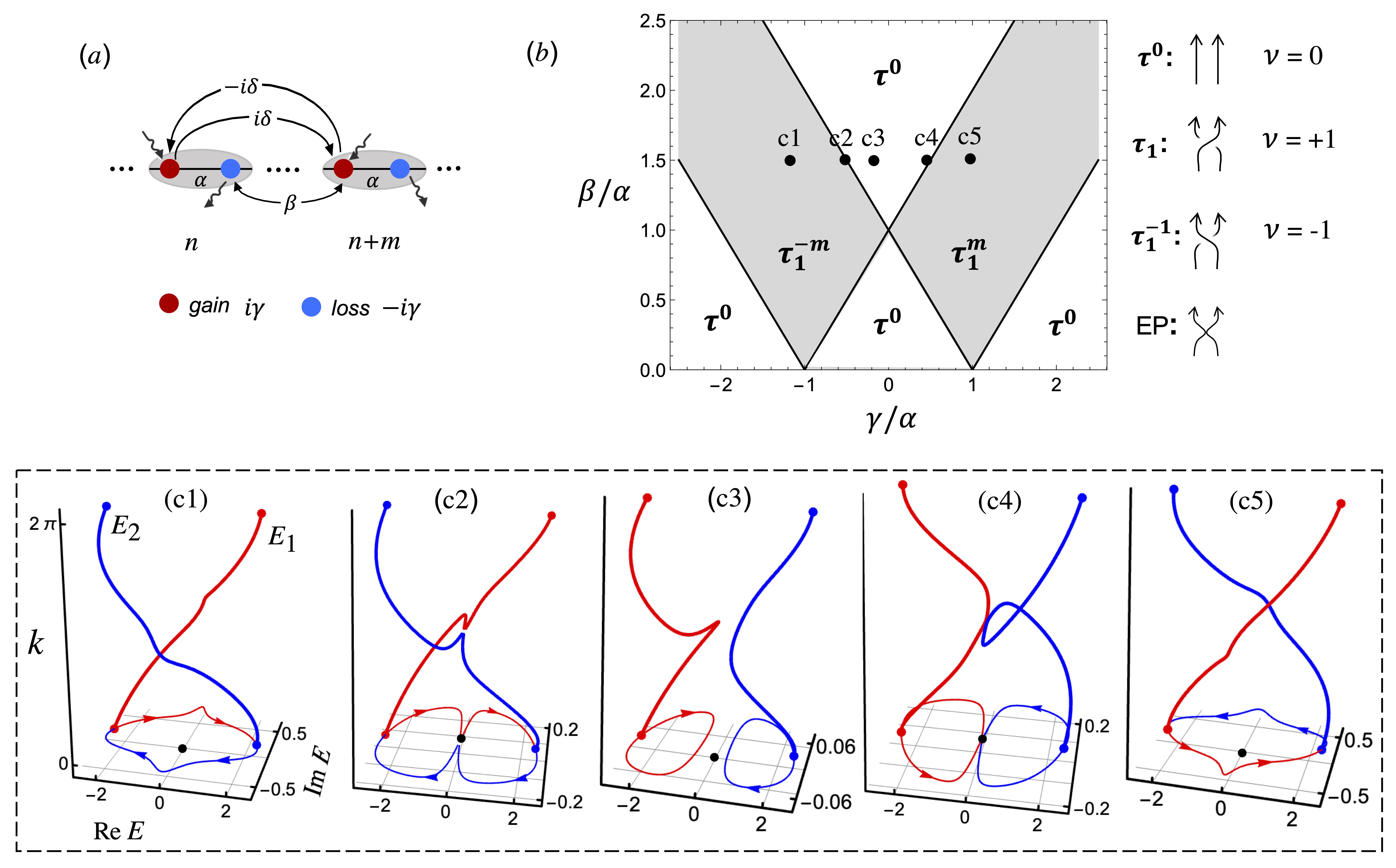} }
 	\caption{(a) A 1D lattice of non-Hermitian dimers for two-band braids. (b) A phase diagram showing different regimes for topologically distinct braids; phase boundaries are the lines of EPs. Topological indices for $\tau_1^{\pm m}$ and $\tau^0$ are $\nu=\pm m$ and $0$, respectively. (c1,c3,c5) Braids $\tau_1^{-1}$, $\tau^0$ and $\tau_1$, respectively for two bands ($E_1$ in red and $E_2$ in blue) as a function of $k$ in $[0,2\pi]$. (c2,c4) Bands evolution in the presence of an EP. The EP (shown by the black dot) occurs at $k=\pi$ and $E_1=E_2=0$. Figs (c1-c5) are obtained by varying $\gamma=-1,-0.5,-0.2,0.5,1$ respectively, and fixed $\beta=1.5$, $\delta=0.3$, $\alpha=1, m=1$. } \label{fig-1}
 \end{figure*}

\begin{figure}[b!]
	\centering{
		\includegraphics[width=0.45\textwidth]{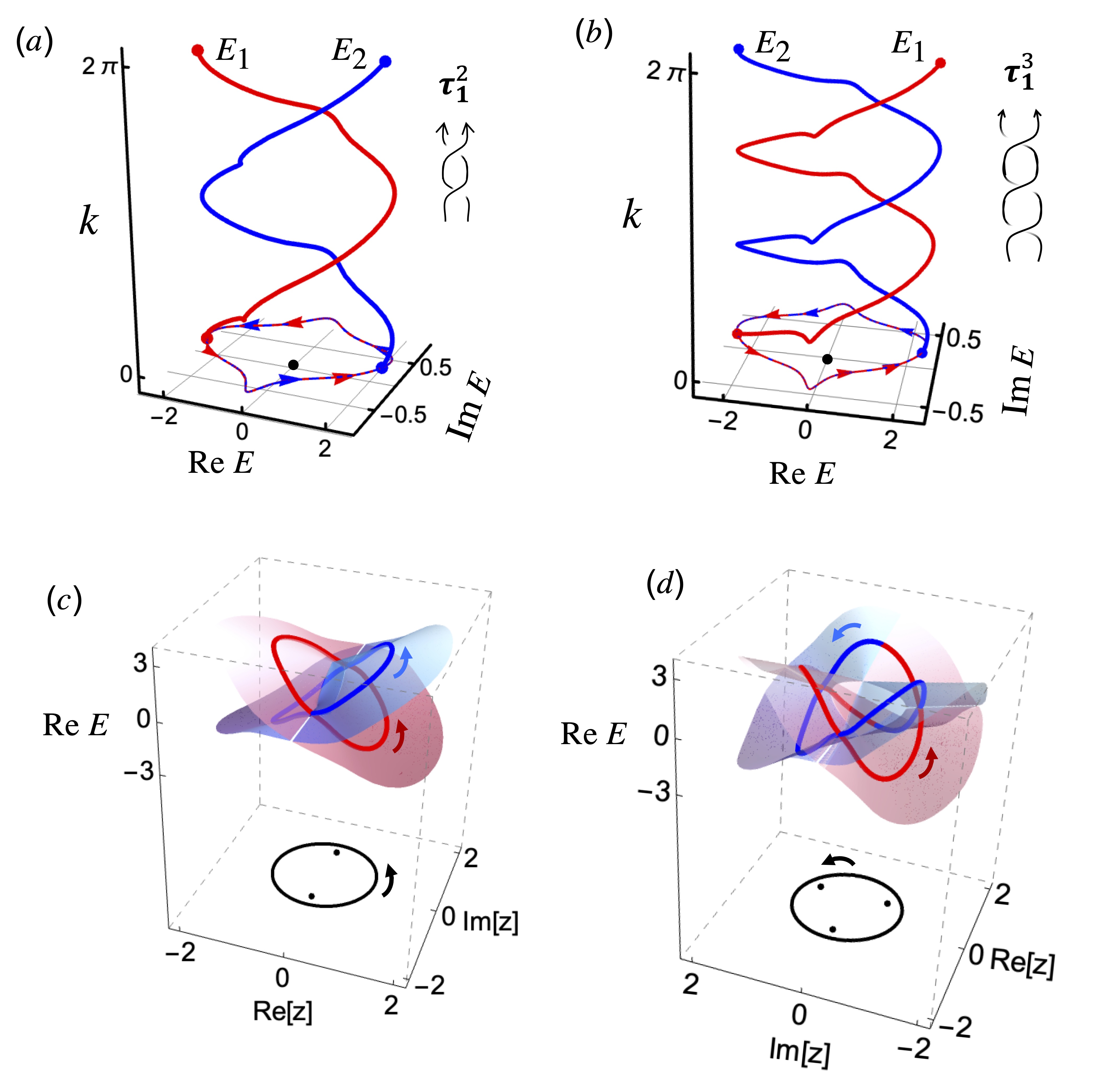} }
	\caption{Higher-order two-band braids (a) $\tau_1^2$ (Hopf link) and (b) $\tau_1^3$ (trefoil knot) when $m=2$ and $3$, respectively. (c,d) Energy loops on the Riemann sheets of $H(z)$ are shown, corresponding to the braids in (a) and (b), respectively. The location of EPs (black dots) is shown inside the first BZ (black circle) on the $z$-plane. Here $\beta=1.5,\gamma=1,\delta=0.3,\alpha=1$.} \label{fig-2}
\end{figure} 

 \begin{figure*}[htb!]
	\centering{
		\includegraphics[width=0.77\textwidth]{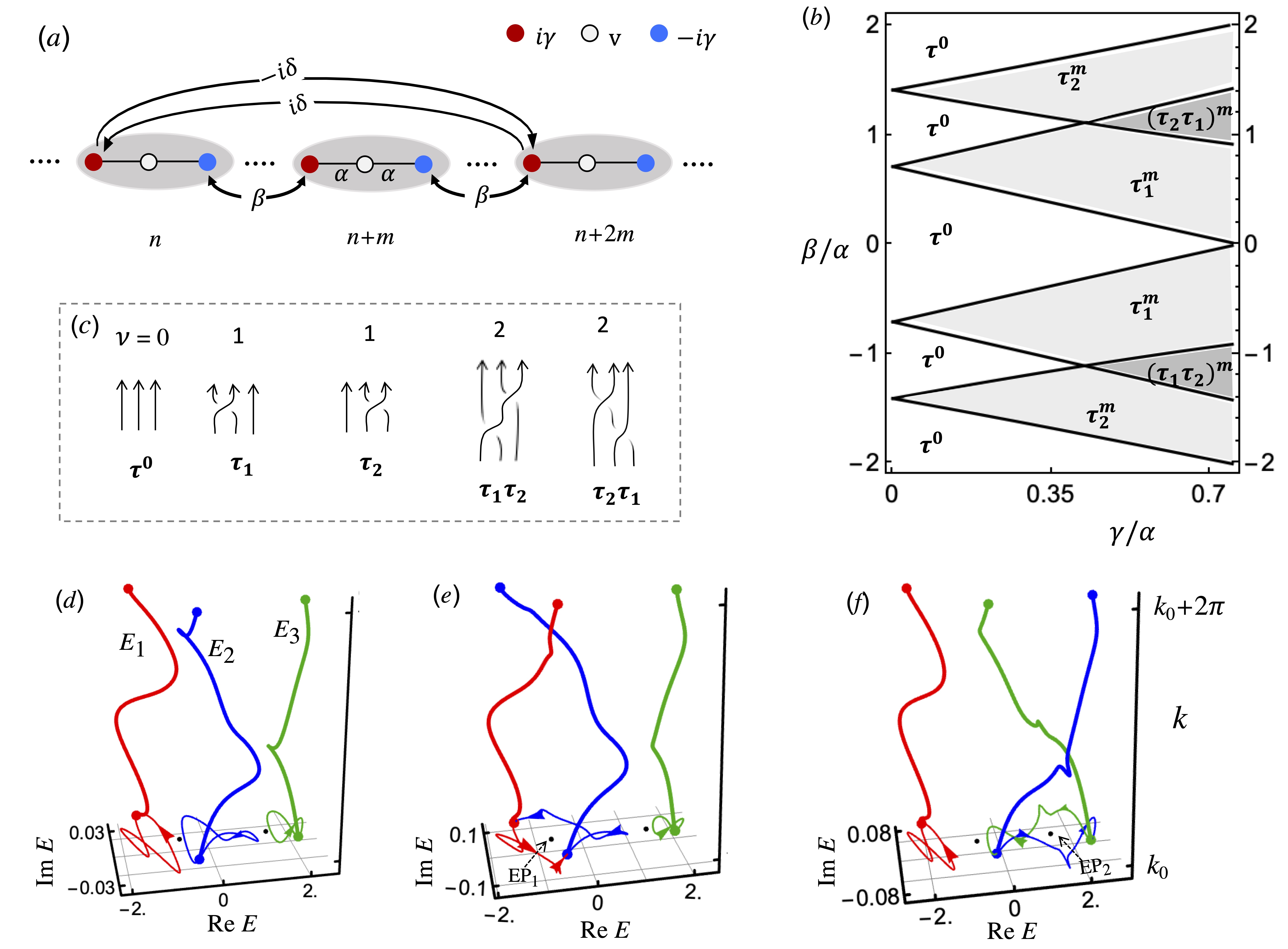} }
	\caption{(a) A lattice of non-Hermitian trimers considered for three-band braids. (b) Phase diagram to realize  identity braid $\tau^0$, $\mathbb{B}_2$ generators $\tau_1$, $\tau_2$, braid words $\tau_1\tau_2$, $\tau_2\tau_1$, and their $m$-fold generalizations are shown. Different braid regimes are separated by EP lines. White, light gray and deep gray regions correspond to the topological index $\nu=0,m$ and $2m$ respectively. (c) Schematic diagrams of a few braids for $m=1$. Braiding of $E_1$ (red), $E_2$ (blue) and $E_3$ (green) bands is shown for (d) $\tau^0$, (e) $\tau_1$ and (f) $\tau_2$.  Figs. (d,e,f) are obtained for $(\beta,\gamma)=(1,0.1), (0.8,0.2), (1.6,0.3)$ respectively, and fixed values of v$=0.7$, $\delta=0.3, \alpha=1, k_0=\pi/4$, $m=1$. Energies at  $(\mbox{EP}_1,\mbox{EP}_2)\simeq(-0.7,1.4)$.} \label{fig-3}
\end{figure*}

 {\it Gain-loss-induced braid group $\mathbb{B}_2$.} 
 A two-band non-Hermitian lattice is considered with  $\alpha$ and $(\beta, \delta)$ being intra-dimer and $m$th nearest-neighbor Hermitian hopping amplitudes, respectively, and $\pm i\gamma$ representing onsite imaginary potentials~[Fig.~\ref{fig-1}a]. The effective-Hamiltonian under periodic boundary condition is given by
 \begin{equation}
 	H(k)=\begin{pmatrix}
 		2 \delta \sin m k+i \gamma  &  \alpha+\beta e^{-i m k}\\
 		\alpha+\beta e^{i m k} & -i\gamma
 	\end{pmatrix},
 	\quad
 \end{equation}
 with  two energy bands
 \begin{equation}\begin{array}{ll}
 		\hspace{-1em}	E_{1,2}(k)=\delta \sin mk\\
 		\quad \mp \sqrt{\alpha^2+\beta^2+2\alpha\beta \cos mk+ (i\gamma +2\delta \sin mk)^2}. 
 	\end{array}
 \end{equation}  
In general, $E_1$ and $E_2$ either braid $m$ times or do not braid at all.  Energy bands permute non-trivially $(E_1,E_2)\rightarrow(E_2,E_1)$ when $m$ is odd, and  trivial permutation $(E_1,E_2)\rightarrow(E_1,E_2)$ occurs for even $m$. Examples of $\tau^0$ and $\mathbb{B}_2$ generators $\tau_1^{\pm1}$, corresponding to unlink and unknot, respectively, are shown in Fig.~\ref{fig-1}c. Higher-order braids $\tau_1^2=\tau_1\tau_1$ (Hopf link) and $\tau_1^3=\tau_1\tau_1\tau_1$ (trefoil knot) are shown in Fig.~\ref{fig-2}.  At an EP, two bands do not braid but  undergo a braid phase transition. EPs are obtained here at the vanishing discriminant of the characteristic polynomial $\mbox{det}(H(k)-E)$, and are given by the lines
\begin{equation}\begin{array}{ll}
		\gamma=\pm(\beta-\alpha)  \quad \mbox{at} \quad k=\pi/m,\\
		\gamma=\pm(\beta+\alpha) \quad\mbox{at} \quad k=0,
	\end{array}\label{eq-3}
\end{equation} 
for $\delta\ne0$. At EPs, both the energies $E_1$ and $E_2$ are identically zero. 

Different braid diagrams are distinguished by the topological integer invariant $\nu\in \mathbb{Z}$ defined by the spectral winding number with respect to the energy at an EP \cite{Ding2022}
 \begin{equation}
 	\nu=\int\limits_{k\in BZ}\frac{dk}{2\pi i} \frac{d}{dk} \ln \det \left(H(k)-\mbox{EP}\right). \label{eq-4}
 \end{equation}
 The braid degree $\nu$ describes how many times the two bands wind around an EP, with the sign indicating the handedness of braids. We obtain $\nu\ne0~(=0)$ for a non-trivial (trivial) braid with $\nu(\tau_1^{\pm m})=\pm m$.  Phase diagram in the  $(\beta,\gamma)$ plane for braids with distinct  $\nu$ is shown in Fig.~\ref{fig-1}b. Different phases are separated by the  exceptional lines given in \eqref{eq-3}. While zero gain loss always results in a trivial braid, a non-zero critical gain loss is required to induce non-trivial braids. Nontrivial braids are also absent whenever the long-range coupling, either $\delta$ or $\beta$, is zero. For a fixed non-zero value of $\beta$, the transition from trivial to non-trivial braid $\tau^0\rightarrow\tau_1^m$ occurs solely by the tuning of onsite gain or loss in the system [Fig.~\ref{fig-1}b]. The braiding direction can be reversed $(\tau_1^m\rightarrow\tau_1^{-m})$ by switching gain into loss $(\gamma\rightarrow -\gamma)$ as a consequence of the non-Hermitian symmetry $H^\dag(k;\gamma)=H(k;-\gamma)$. Furthermore, braids with $m$ turns are obtained by long-range coupling between dimers and their $m$-th nearest-neighbors [Fig.~\ref{fig-1}a].

To further understand the role of EPs on the formation of braids and knots we obtain the energy Riemann surfaces of $H(z)$ where $z\rightarrow r e^{i k}$ and $r\ge0$. The first BZ is now a circle on the complex $z$-plane. In general, for a $m$-th order braid or knot, the two energy bands either encircle in the vicinity of a single EP $m$-times or encircle $m$ distinct EPs once inside the first BZ. Locations of EPs on the $z$-plane are given by 
\begin{equation}
	z_{\tiny{\mbox{EP}}}=\left(\frac{\alpha^2+\beta^2-\gamma^2}{2\alpha \beta}\right)^{1/m} e^{i\frac{(2j-1)\pi}{m}}, \quad j=1\cdots m,
\end{equation}
with $E_{1,2}(z_{\tiny{\mbox{EP}}})=0$. Examples of Hopf link and trefoil knot on energy Riemann surfaces of $H(z)$ are shown in Fig.~\ref{fig-2} with two and three EPs, respectively, lying inside the first BZ. 
 
\begin{figure}[tb!]
	\centering{
		\includegraphics[width=0.49\textwidth]{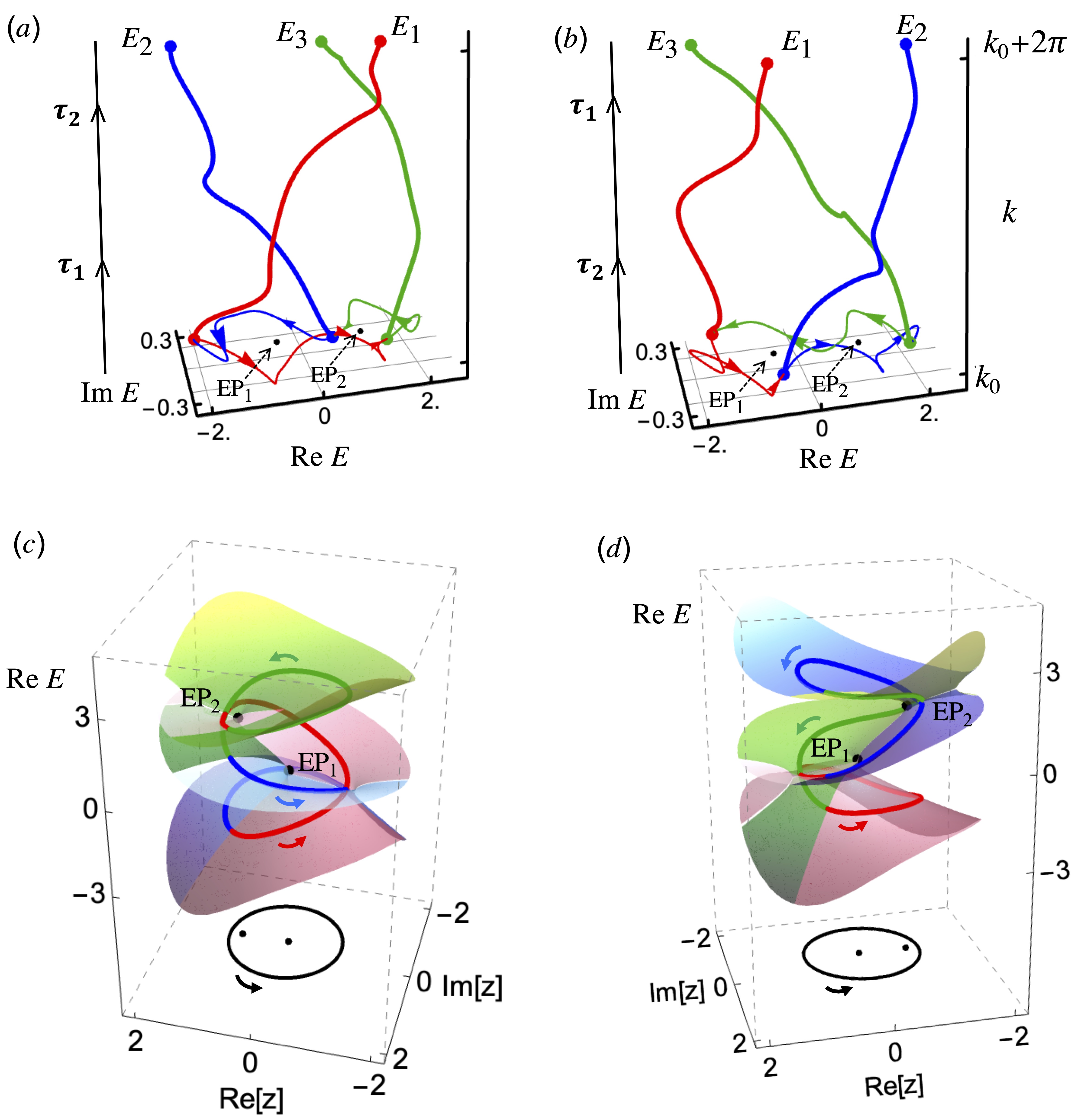} }
	\caption{Braids (a) $\tau_1\tau_2$ and (b) $\tau_2\tau_1$ show non-Abelian energy swapping. (c,d) Corresponding non-Abelian loops are shown on the energy Riemann surfaces of $H(z)$ in the vicinity of two EPs  $(\mbox{EP}_1,\mbox{EP}_2)\simeq(-0.51,1.15)$ lying inside the first BZ.  The $\tau_1\tau_2$ loop encircles first EP$_1$, then EP$_2$, and the $\tau_2\tau_1$ loop encircles first EP$_2$, then EP$_1$ when the BZ is traversed in anticlockwise.  Here $\delta=0.3,\mbox{v}=0.7, \gamma=0.7, \alpha=1$, $k_0=\pi/4$ and $m=1$, but $\beta=-1.2$ in (a,c) and $1.2$ in (b,d) respectively. } \label{fig-4}
\end{figure}

{\it Gain-loss induced non-Abelian braid group $\mathbb{B}_3$.} We now consider a minimal model of non-Hermitian lattice of trimers with Bloch Hamiltonian given by 
\begin{equation}
	H(k)=\begin{pmatrix}
		-2 \delta \sin 2 m k+i \gamma  & \alpha & \beta e^{-i m k}\\
		\alpha & \mbox{v} & \alpha\\
		\beta e^{i  m k} &\alpha & -i\gamma
	\end{pmatrix}.\label{Eq-6}
\quad
\end{equation}
Here $\alpha$ is the intra-cell coupling, $\beta$ is the $m$-th, and $\delta$ is the $2m$-th nearest-neighbor Hermitian coupling [Fig.~\ref{fig-3}a]. $\pm i\gamma$ and $\mbox{v}$ are imaginary and real onsite potentials respectively. For the sake of brevity below, we consider braids corresponding to $m=1$. Three energy bands $E_1$, $E_2$, and $E_3$, as a function of $k$, constitute  different braid elements for the group $\mathbb{B}_3$, provided the discriminant of the characteristic polynomial is non-zero to avoid degeneracy. The phase diagram in the $(\beta,\gamma)$ plane for a few lowest-order braids $\tau^0,\tau_1,\tau_2,\tau_1\tau_2$ and $\tau_2\tau_1$ is computed and shown in Fig.~\ref{fig-3}. Phase boundaries, i.e., exceptional lines between two distinct braids, correspond to the vanishing discriminant of the cubic polynomial $\mbox{det}(H-E)$. Numerically computed examples of braid generators $\tau_1$, $\tau_2$, and identity braid $\tau^0$ are shown in Fig.~\ref{fig-3}, with energy permutation $(E_1,E_2,E_3)\rightarrow (E_2,E_1,E_3)$, $(E_1,E_3,E_2)$ and $(E_1,E_2,E_3)$, respectively. Phase transition $\tau^0 \rightarrow \tau_1$ occurs through the coalescence of bands $E_1$ and $E_2$ at $\mbox{EP}_1$, and $\tau^0\rightarrow \tau_2$ occurs at the $\mbox{EP}_2$, through the coalescence of $E_2$ and $E_3$. The determination of topological index for braids (and knots) for more than two bands is a technically challenging task and is usually done by polynomial invariant \cite{Kauffman2013,Hu2021,Kassel2008}. Here we adopt the global winding number approach, which correctly captures the essential braid invariance properties of interest. The  index is given by 
$\nu=\sum\limits_j \nu_j,$
where $\nu_j$ is the winding number around the $j$-th exceptional point $\mbox{EP}_j$ [refer to Eq.~\eqref{eq-4}]. In general, $\nu=P-Q$, where $P$ is the number of generators and $Q$ is the number of inverse generators in a braid word. In the absence of exact expressions, we numerically obtain EPs for the examples given in Fig.~\ref{fig-3}. Both the EPs lie outside the energy loops for the identity braid $\tau^0$, but EP$_1$ lies inside the $E_1-E_2$ loop of $\tau_1$, and EP$_2$ lies inside the $E_2-E_3$ loop of $\tau_2$.  Correspondingly, we obtain $\nu(\tau^0)=0$ and $\nu(\tau_1)=\nu(\tau_2)=1$.  

One of the remarkable features of three-bands braiding is that the concatenation of two braids is non-Abelian. For example, $\tau_1\tau_2\ne\tau_2\tau_1$ as the braid diagrams for these two products are different. (Particular examples are shown in Fig.~\ref{fig-4} together with the associated energy loop on three Riemann sheets). For the examples in Fig. 4, two EPs lie inside the energy loops. The spectral winding number is given by $\nu(\tau_1\tau_2)=\nu(\tau_2\tau_1)=2$. Inequivalence of  $\tau_1\tau_2$ and $\tau_2\tau_1$, however, stems from the collective behavior of multiple EP singularities and universal non-Abelian energy evolution around them (see Refs.~\cite{Zhong2018,Pap2018,Guo2023}). Note the different orders in which two EPs are encircled for the two braids: anticlockwise loop $\tau_1\tau_2$ encircles EP$_1$ followed by EP$_2$, the order is reversed for  $\tau_2\tau_1$ (see Fig.~\ref{fig-4}). Non-Abelian braiding has an  important observable effect in terms of energy permutations; one obtains $(E_1,E_2,E_3)\rightarrow (E_2,E_3,E_1)$ for $\tau_1\tau_2$ and $(E_1,E_2,E_3)\rightarrow(E_3,E_1,E_2)$ for $\tau_2\tau_1$ [Fig.~\ref{fig-4}(a,b)].  For a fixed gain-loss potential, the two non-Abelian braids can be realized by changing the sign of the coupling $\beta$ (see phase diagram in Fig.~\ref{fig-3}b). Note that the engineering of negative coupling was demonstrated in Refs.~\cite{Keil2016,Zhang2021}. 

 Similar to the two band cases, the handedness of  generators of $\mathbb{B}_3$ can also be reversed i.e. $(\tau_{1},\tau_2)\rightarrow(\tau_{1}^{-1},\tau_2^{-1})$ by $\gamma\rightarrow -\gamma$. This can be understood from the relation $E(k;-\gamma)=E(k;\gamma)^*$ as a consequence of $H(k;-\gamma)=H^\dag(k;\gamma)$.  In addition, $m$-fold generalization of all the non-trivial braids discussed above, i.e., braids $\tau_{1}^m, \tau_2^m, (\tau_1\tau_2)^m$, $(\tau_2\tau_1)^m$ can be generated by considering long-range coupling $\pm i\delta$ between gain sites of $(2m)$-th nearest-neighbor unit cells and coupling $\beta$ between loss and gain sites of $m$-th nearest-neighbor cells (Fig.~\ref{fig-3}a). Interestingly, the phase diagram in Fig. 3b remains intact for arbitrary $m$. A few numerical examples of higher-order braids,  inverses of generators and their concatenation are given in the supplementary material. 

Looking ahead, the effect of quantum noise on spectral braids, interplay of non-Abelian `eigenstate' braiding, Wilczek–Zee
connection \cite{Park2022}, and open-boundary induced non-Hermitian skin-effect \cite{Zhang2022}, kept outside the scope of current investigation, merit further exploration in NH multiband lattices.  From an  application point of view, the proposed braiding and non-Abelian swapping of energy bands open pathways to designing non-Hermitian fault-tolerant topological computing devices where gain and loss are generated by controllable interaction with the environment   \cite{Ota2020}.

\newpage

\setcounter{figure}{0}
\renewcommand{\thefigure}{S\arabic{figure}}

\begin{widetext}
	
\begin{center}
	 {\large Supplementary Material}
 \end{center}

\bigskip

 Here we provide few numerical examples of braid words consisting inverse generators $\tau_1^{-1}$, $\tau_2^{-1}$, and examples of higher-order generalizations of three-band braids, obtained by the prescriptions described in the main text.
 
\begin{figure*}[h!]
	\centering{
		\includegraphics[width=0.7\textwidth]{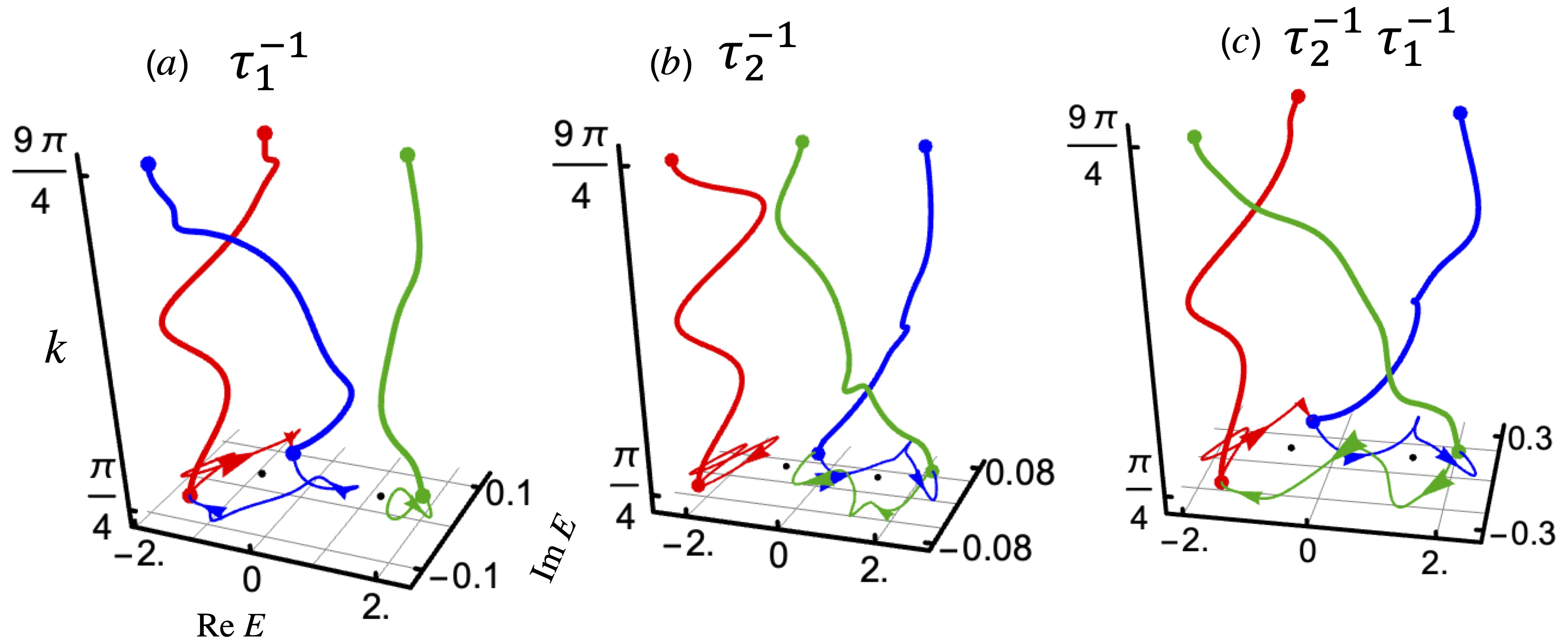} }
	\caption{ Examples of inverse of braid generators and their concatenation obtained for (a) $\gamma=-0.2$, (b) $\gamma=-0.3$, (c) $\gamma=-0.7$. Other parameters remain unchanged as in Fig. 3(e), 3(f), and 4(b), respectively, of the main text. The topological indices for these braids are $\nu(\tau_1^{-1})= \nu(\tau_2^{-1})=-1, \nu({\tau_2^{-1}\tau_1^{-1}})=-2$. The handedness of all the energy knots in the (Re~$E$,Im~$E$) plane for braids $\tau^{-1}$, $\tau_2^{-1}$, and $\tau_2^{-1}\tau_1^{-1}$ is opposite compared to  that of $\tau_1$, $\tau_2$, and $\tau_2\tau_1$, respectively, shown in Figs. 3(e), 3(f) and 4(b) of the main text. But corresponding band permutations are unaffected.  } \label{fig-s1}
\end{figure*}

\begin{figure*}[b!]
	\centering{
		\includegraphics[width=0.7\textwidth]{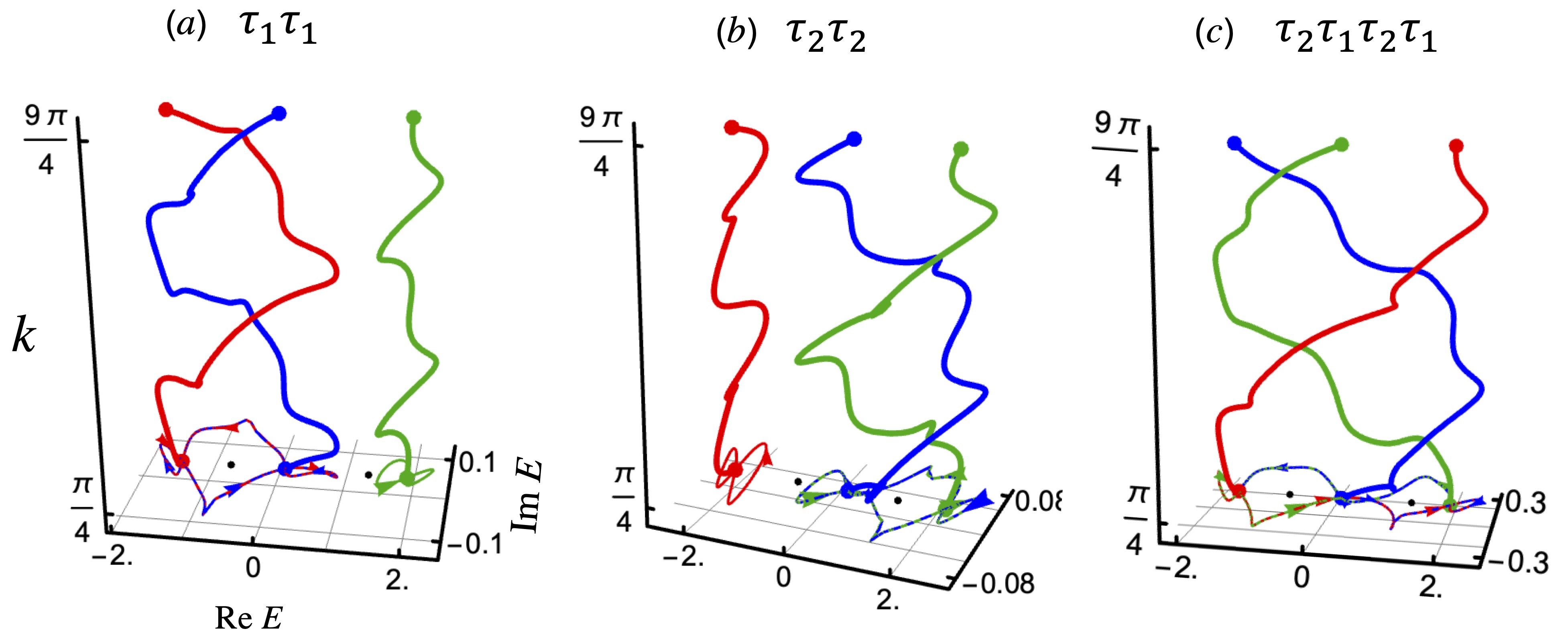} }
	\caption{ Examples of higher order generalizations of braids consisting three energy bands. All the parameters, except $m$, remain same as in Fig. 3(e), 3(f) and 4(b), respectively, of the main text.  Here $m=2$. The topological numbers for these braids are $\nu(\tau_1\tau_1)=2, \nu(\tau_2\tau_2)=2, \nu(\tau_2\tau_1\tau_2\tau_1)=4$. Braids in (a) and (b) correspond to trivial band permutation, whereas braid in (c) corresponds to $(E_1,E_2,E_3 )\rightarrow(E_3,E_1,E_2)$.} \label{fig-s2}
\end{figure*}

\end{widetext}


	\end{document}